\documentclass[conference]{IEEEtran}
\usepackage{cite}
\usepackage{amsmath,amssymb,amsfonts}
\usepackage{algorithmic}
\usepackage{graphicx}
\usepackage{textcomp}
\usepackage{xcolor}
\usepackage{subfig}
\usepackage{makecell}
\usepackage{booktabs}
\usepackage{subfig}
\usepackage{tabularx}

\usepackage{pgfplots}
\usepgfplotslibrary{groupplots,dateplot}
\usetikzlibrary{patterns,shapes.arrows}
\pgfplotsset{compat=newest}

\def\BibTeX{{\rm B\kern-.05em{\sc i\kern-.025em b}\kern-.08em
    T\kern-.1667em\lower.7ex\hbox{E}\kern-.125emX}}

\newcolumntype{Y}{>{\centering\arraybackslash}X}

\makeatletter
\def\footnoterule{\kern-3\p@
  \hrule \@width 2in \kern 2.6\p@} 
\makeatother

\begin{document}

\title{Divide and Save: Splitting Workload\\ Among Containers in an Edge Device \\ to Save Energy and Time\\

\thanks{This work has been supported by the EU H2020 MSCA ITN project Greenedge (grant no. 953775), and by the EU under the Italian National Recovery and Resilience Plan (NRRP) of NextGenerationEU, partnership on “Telecommunications of the Future” (PE0000001 - program “RESTART”).}
}

\author{\IEEEauthorblockN{Aria Khoshsirat, Giovanni Perin, and Michele Rossi}
\IEEEauthorblockA{\textit{Department of Information Engineering (DEI)} \\
\textit{University of Padova (Padova, Italy)}\\
Emails: aria.khoshsirat@unipd.it, giovanni.perin.1@unipd.it, michele.rossi@unipd.it}
}

\IEEEoverridecommandlockouts

\maketitle
\makeatletter
\def\ps@IEEEtitlepagestyle{%
  \def\@oddfoot{\mycopyrightnotice}%
  \def\@oddhead{\hbox{}\@IEEEheaderstyle\leftmark\hfil\thepage}\relax
  \def\@evenhead{\@IEEEheaderstyle\thepage\hfil\leftmark\hbox{}}\relax
  \def\@evenfoot{}%
}
\def\mycopyrightnotice{%
  \begin{minipage}{\textwidth}
  \centering \scriptsize
  \copyright~2023 IEEE. Personal use of this material is permitted. Permission from IEEE must be obtained for all other uses, in any current or future media, including reprinting/republishing this material for advertising or promotional purposes, creating new collective works, for resale or redistribution to servers or lists, or reuse of any copyrighted component of this work in other works.
  \end{minipage}
}
\makeatother

\begin{abstract}

The increasing demand for edge computing is leading to a rise in energy consumption from edge devices, which can have significant environmental and financial implications. To address this, in this paper we present a novel method to enhance the energy efficiency while speeding up computations by distributing the workload among multiple containers in an edge device. Experiments are conducted on two Nvidia Jetson edge boards, the TX2 and the AGX Orin, exploring how using a different number of containers can affect the energy consumption and the computational time for an inference task. To demonstrate the effectiveness of our splitting approach, a video object detection task is conducted using an embedded version of the state-of-the-art YOLO algorithm, quantifying the energy and the time savings achieved compared to doing the computations on a single container. The proposed method can help mitigate the environmental and economic consequences of high energy consumption in edge computing, by providing a more sustainable approach to managing the workload of edge devices.
\end{abstract}


\section{Introduction}
Multi-access edge computing (MEC) is a rapidly growing field that entails performing computational tasks at the network's edge, closer to the data source. This approach, other than inherently lowering the total communication and computation latency, provides other benefits, such as improving security and privacy. However, as the demand for edge computing services increases, it becomes increasingly important to address the issue of the joint optimization of energy efficiency and computational time. In this paper, we refer to energy efficiency as to the ability of executing a computational unit of a task (CPU cycles) using the least amount of energy possible. This is crucial for two main reasons. First, edge computing devices are often powered by batteries or have limited power supplies. Energy efficiency is thus key to ensure their operation for long periods of time and for making their computations feasible, even with the limited amount of energy available. Second, as the number of edge devices increases, the overall energy consumption of the edge computing system increases accordingly. This can lead to a significant environmental footprint and to unsustainable economic costs for the service providers.

In this paper, we investigate how splitting a computing task into multiple containers within the same Nvidia Jetson device impacts the energy drained and the execution time. Instances of the popular deep learning-based algorithm ``You only look once'' (YOLO)~\cite{yolo} are used, employing YOLOv4-tiny~\cite{yolov4Tiny}, which is suitable for execution on energy and memory constrained hardware. According to a recent survey~\cite{oleghe2021container}, MEC schedulers decide, other than container placement and migration, \emph{how} incoming tasks should be allocated to containers. This paper provides useful insights on how to optimally allocate \emph{splittable tasks} into multiple containers from an energy and execution time perspective. As such, it is indirectly related to the concept of \emph{split computing}~\cite{matsubara2022split}, i.e., the division of a neural network (NN) model into a head and a tail, to be executed on different devices (and, thus, on different containers). Notably, however, split-computing introduces time dependencies, as the head has to be executed {\it before} the tail. The application chosen in this paper, namely, YOLO, does not require keeping temporal dependencies into account since frames are processed {\it independently} of one another. For now, we leave the problem of time dependencies for future studies.

\noindent To summarize, the main contributions of this work are:
\begin{itemize}
    \item We provide a new approach to improve energy efficiency and decrease the computational time of edge computing by splitting the computations among multiple containers.
    \item Testbed results of our experiments are shown, performed on two commercial edge devices of the series Nvidia Jetson, namely, the TX2 and the AGX Orin. They provide evidence of the effectiveness of our method in reducing both energy consumption and processing~time.
    \item We obtain simple convex models for the energy consumption and inference time of commercial edge devices as a function of the number of containers used to split the task. Such models can be used to effectively schedule the computation of workload in a MEC server.
\end{itemize}  
We believe this paper will be helpful for researchers, practitioners, and policymakers who are interested in the energy and time optimization of MEC platforms.

The remainder of this paper is organized as follows: The related work is briefly reviewed in Section~\ref{sec:rel}. In Section~\ref{sec:back}, we present some background on YOLO and Docker containers. The experimental setup is explained in Section~\ref{sec:setup}. The method is detailed in Section~\ref{sec:meth}, while the results are given together with the discussion in Section~\ref{sec:res}. Finally, conclusions and future research lines are discussed in Section~\ref{sec:concl}.

\section{Related Work}
\label{sec:rel}



Recent work has modeled the energy consumption of neural network models over Nvidia Jetson hardware~\cite{rodrigues2017fine, suzen2020benchmark,holly2020profiling,lahmer2022energy}. Specifically, in~\cite{rodrigues2017fine}, the authors develop a framework to measure the energy consumption of specific layers of a convolutional neural network (CNN) on a Jetson TX1. Papers~\cite{suzen2020benchmark} and~\cite{holly2020profiling} instead profile the energy consumption of Jetsons TX2 and Nano, providing an optimized set of parameters to increase energy efficiency. In our previous work~\cite{lahmer2022energy}, we profiled Jetsons TX2 and Xavier, also providing models to estimate their energy consumptions based on the neural network layers features. In~\cite{dockkvm}, the authors compare containers, virtual machines and multi-thread computing on a parallel benchmark. They show that the overhead of using Docker containers is very small compared to virtual machines.

To the best of our knowledge, our paper is the first providing an energy and latency profiling of task splitting on Jetson devices. Notably, this is of great interest for a number of applications concerning container allocation and migration~\cite{oleghe2021container} and energy efficiency~\cite{shalavi2022energy} at the network's edge. In~\cite{kaur2020keids}, the authors develop a Kubernetes-based~\cite{kubernetes} container scheduler that takes into account the problem of energy minimization. A similar problem involving container migration in an urban vehicular context is treated in~\cite{perin2022ease}, with the objective of minimizing the MEC carbon footprint. However, both these papers consider the tasks as unsplittable monolithic entities. In the present work, we show that splitting tasks among multiple containers, when possible, is beneficial both in terms of energy consumption and computational time.

\section{Application Background} 
\label{sec:back}
\subsection{YOLO}

YOLO~\cite{yolo} is a state-of-the-art object detection algorithm that is widely used in computer vision applications. This algorithm is based on a deep CNN architecture and can detect objects within an image or a video frame in \mbox{real-time}. One of the key advantages of YOLO is its ability to detect multiple objects within a single forward pass of the network, as opposed to traditional object detection methods that require multiple runs. This allows the model to achieve a faster detection rate and a higher accuracy compared to previous algorithms. Additionally, YOLO's architecture allows for easy integration with other computer vision tasks such as object tracking and semantic~segmentation. YOLO has shown good performance in various applications such as self-driving cars, surveillance and augmented reality~(AR) systems. However, its performance is sensitive to the quality of the training dataset and the architecture design. Therefore, researchers have proposed improvements and variants to the original algorithm to improve it.
In this paper, as a case study, we consider an object detection task on video data using the YOLOv4-Tiny~\cite{yolov4Tiny} algorithm. YOLOv4-tiny is based on YOLOv4~\cite{Yolov4} with the objective of making the YOLO neural network structure simpler (fewer parameters), which makes it suitable for mobile and embedded devices with power and memory constraints.


\begin{table}
\vspace{0.13cm}
\def\arraystretch{1.5}
\caption{Device hardware specifications}
\begin{center}
\resizebox{\columnwidth}{!}{\begin{tabular}{lcc}
\toprule
 & \textbf{Jetson TX2}& \textbf{Jetson AGX Orin} \\
\midrule\midrule
\textbf{CPU}& \makecell{Quad-core ARM Cortex-A57 \\+ Dual-core Denver 2$^{\mathrm{*}}$} & \makecell{12-core Arm Cortex-A78}  \\
\textbf{GPU}& 256-core NVIDIA Pascal& 2048-core NVIDIA Ampere   \\
\textbf{Memory}& 8 GB 128-bit LPDDR4 & 32GB 256-bit LPDDR5   \\
\textbf{Performance}& 1.33 TFLOPs & 200 TOPS   \\
\bottomrule
\multicolumn{3}
{l}{$^{\mathrm{*}}$Denver cores are turned off by default for consistency}
\end{tabular}}
\label{config}
\end{center}
\vspace{-0.7cm}
\end{table}

\subsection{Docker Containers}
To create containers we have used Docker~\cite{docker}. A Docker container is a lightweight executable package including everything that is needed to run a piece of software, including the code, a runtime, system tools, libraries, and settings. Containers provide a consistent way to package and distribute software, making it easier to deploy and run applications on different environments. Containers are isolated from one another and from the host system, so they can run without conflicts, even when multiple containers run on the same host. Containers are based on Docker images, which are snapshots of a container's file system at a specific point in time. It is possible to limit the CPU resources of a Docker container, which is usually done to control the amount of resources the host uses to execute it. This may be used to enforce a fair distribution of computing resources among the different running containers and processes. The ``{-}{-}cpus'' runtime option allows us to specify the number of CPU cores that the container can use. For instance, the command "docker run {-}{-}cpus=2 Yolo-Container" limits the created container to use only 2 CPU cores.

\begin{figure*}
    \centering
    \subfloat[Time]{\resizebox{.39\textwidth}{!}{\begin{tikzpicture}

\definecolor{darkslategray38}{RGB}{38,38,38}
\definecolor{lightgray204}{RGB}{204,204,204}
\definecolor{mediumorchid}{RGB}{186,85,211}

\begin{axis}[
axis line style={lightgray204},
tick align=outside,
tick pos=left,
x grid style={lightgray204},
xlabel=\textcolor{darkslategray38}{Number of CPU cores},
xmajorgrids,
xmin=0.15, xmax=12.25,
xtick style={color=darkslategray38},
yticklabels={0,0,5,10,15,20},
y grid style={lightgray204},
ylabel=\textcolor{darkslategray38}{Inference Time ($\times 10^2$ s)},
ymajorgrids,
ymin=0, ymax=2000,
ytick style={color=darkslategray38}
]
\addplot [only marks, ultra thick, red, mark=o, mark size= 3pt]
table {%
0.8 1923.818377
1 1428.508946
2 537.845278
3 340.478856
4 335.602946
};
\addlegendentry{TX2 data}
\addplot[domain=0.6:4, thick, dashed] {6518.65 * exp{-1.765 * x + 326.77}};
\addlegendentry{TX2 fit}

\addplot [only marks, ultra thick, blue, mark=triangle, mark size= 3 pt]
table {%
0.4 624.392021
1 186.254587
2 63.098788
3 62.881196
4 50.273535
5 49.766862
6 49.88659
8 51.143005
10 51.438918
12 51.665154
};
\addlegendentry{Orin data}
\addplot[domain=0.1:12, dotted, thick] {1510.64 * exp{-2.432 * x + 53.35}}; 
\addlegendentry{Orin fit}
\end{axis}

\end{tikzpicture}}}
    \subfloat[Energy]{\resizebox{.38\textwidth}{!}{\begin{tikzpicture}

\definecolor{darkslategray38}{RGB}{38,38,38}
\definecolor{lightgray204}{RGB}{204,204,204}
\definecolor{mediumorchid}{RGB}{186,85,211}

\begin{axis}[
axis line style={lightgray204},
tick align=outside,
tick pos=left,
x grid style={lightgray204},
xlabel=\textcolor{darkslategray38}{Number of CPU cores},
xmajorgrids,
xmin=0.15, xmax=12.25,
xtick style={color=darkslategray38},
yticklabels={0,0,1,2,3,4,5},
y grid style={lightgray204},
ylabel=\textcolor{darkslategray38}{Energy Consumption (kJ)},
ymajorgrids,
ymin=500, ymax=5500,
ytick style={color=darkslategray38}
]
\addplot [only marks, ultra thick, red, mark=o, mark size= 3pt]
table {%
0.8 3268.84162865821
1 2539.47271842151
2 1233.05013960583
3 949.307547479176
4 943.936265208808
};
\addlegendentry{TX2 data}
\addplot[domain=0.6:4, thick, dashed] {9714.04 * exp{-1.787 * x + 932.72}};
\addlegendentry{TX2 fit}

\addplot [only marks, ultra thick, blue, mark=triangle, mark size= 3 pt]
table {%
0.4 5488.99696061117
1 1665.88044546993
2 664.527237501471
3 768.040256780007
4 665.475100540928
5 659.437109104965
6 658.498957712066
8 673.714665199206
10 676.928369695454
12 679.441260977017
};
\addlegendentry{Orin data}
\addplot[domain=0.1:12, dotted, thick] {13924.82 * exp{-2.653 * x + 672.86}}; 
\addlegendentry{Orin fit}
\end{axis}

\end{tikzpicture}}}
    \caption{Plots showing the variation in inference time and energy consumption as the number of CPU cores allocated to a single container is increased, on Jetson TX2 and Jetson AGX Orin edge devices.}
    \label{Tx2contain}
\end{figure*}
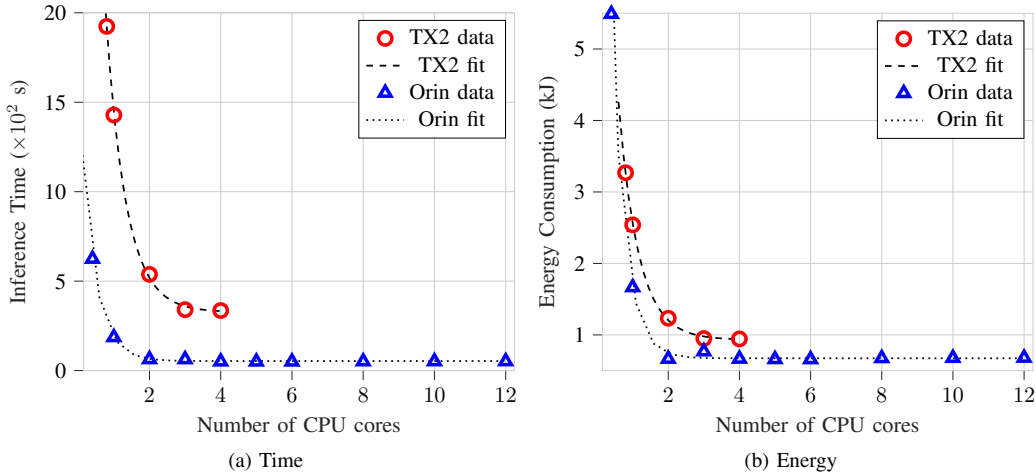

\section{Experimental Setup}
\label{sec:setup}
For the experiments of this work, we employed two Nvidia edge devices: Jetson TX2 and Jetson AGX~Orin. The Jetson TX2 is a previous generation cost-effective edge device with a powerful GPU and CPU, which can run complex algorithms and process large amounts of data at the edge. The Jetson AGX Orin, on the other hand, is a newer and more powerful platform with a higher number of CPU cores and more memory, making it ideal for running more demanding tasks. Both platforms support a wide range of neural networks and machine learning frameworks, making them versatile tools for research and industry applications alike. Table~\ref{config} shows the hardware specifications of these devices. Note that for our experiments we have only used 4 ARM CPU cores on the Jetson TX2 since the Denver cores are disabled by default in the device due to incompatible~performance.

Although these devices have powerful GPUs, we have used only the CPU for our experiments. This is because CPU cores can be readily divided among containers, whereas a container's allocation in terms of GPU resources cannot be easily enforced. In fact, without any restriction on GPU usage, containers end up competing for the majority of GPU resources, making it challenging for the GPU scheduler to operate efficiently. For more than two containers on the TX2 and four containers on the AGX Orin, this leads to irregular and slow computation. It is worth noting that in an edge device that has multiple GPUs, the method outlined in this paper can still be applied by assigning each GPU to a specific (different) container, with no conflicts.

For measuring the power of these devices, we have used the \mbox{built-in} power monitoring sensors included in both Nvidia boards. Such sensor can be read with a sampling time of about 10 milliseconds, which is accurate enough for our experiments. The energy consumption is then calculated by taking the sum of the power readings multiplied by the time period between subsequent power samples.

The base experiment that is evaluated in this paper consists of the application of object detection on a 30-second-long video, using YOLOv4-Tiny. Notably, through experimentation with object detection on videos with differing formats, we found that the {\it number of frames} in a video has the greatest impact on the energy and time needed for YOLO inference. 
Other characteristics of a video, such as the frame size, the bitrate, or even the number of objects per frame, have minimal effect on the time and energy, and thus can be neglected.



As an initial step to demonstrate a baseline, we performed the task on each device using one container with a varying number of CPU cores available to it. Fig.~\ref{Tx2contain} shows the time and energy of performing the object detection task for the full \mbox{30-second} video, as a function of the number of CPUs allotted to the container. This is a real number and in our experiment is varied from 0.1 (ten percent of one CPU core) to the number of cores available on the device. On the Jetson TX2, we observe that using four CPU cores only results in a slight improvement in time and energy efficiency when compared to using three cores. Doing the same experiment on the AGX Orin device exhibits a similar behavior; employing more than two CPU cores minimizes efficiency. This led us to conjecture that increasing the utilization of the CPU cores for computations and parallelizing the workload, could enhance the efficiency of the computations on edge devices.



\section{Methodology}
\label{sec:meth}
This study aims to propose a method to decrease the energy consumption and computational time of an inference task, namely, the object detection task introduced in the previous section, on edge devices. To achieve this, the following methods are used:

\begin{enumerate}
    \item \textbf{Data splitting:} The test data, in our case the whole input video, is split into equal size segments. The splitting is done along the time dimension of the video, resulting in the same number of frames for each segment, to be used as input for the object detection task.
    \item \textbf{Creating containers:} We subsequently generate a number of containers matching the number of data segments, with each container running an instance of the YOLO model to perform the inference.
    \item \textbf{Dividing computational resources:} The processing units, i.e., the CPU cores, are evenly split among the containers. Each container receives a share of the maximum processing capacity of the device, depending on the number of containers that are created. 
    \item \textbf{Parallelization:} The inference is carried out on all the containers {\it simultaneously}, each accessing its designated segment of input data and using the available processing units. The results from all the containers are then combined and presented to the user.
    
\end{enumerate}

\begin{figure}
\centering
\subfloat{\includegraphics[width=.9\columnwidth]{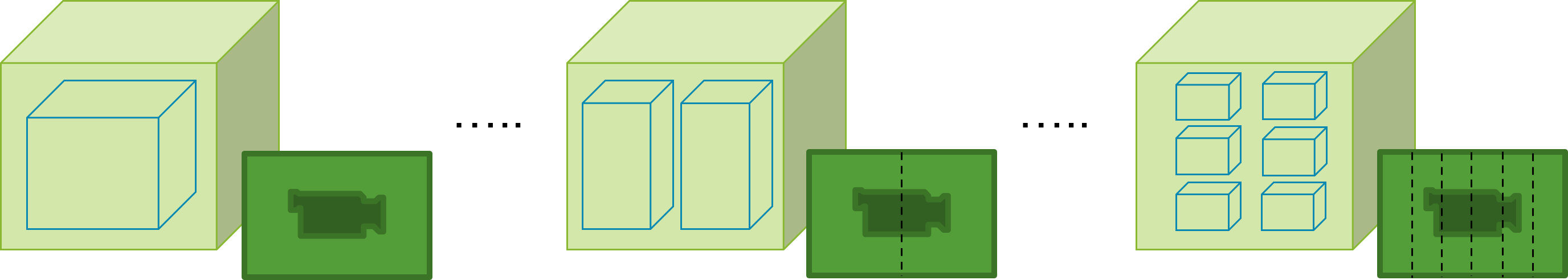}}\\
\subfloat{\includegraphics[width=.9\columnwidth]{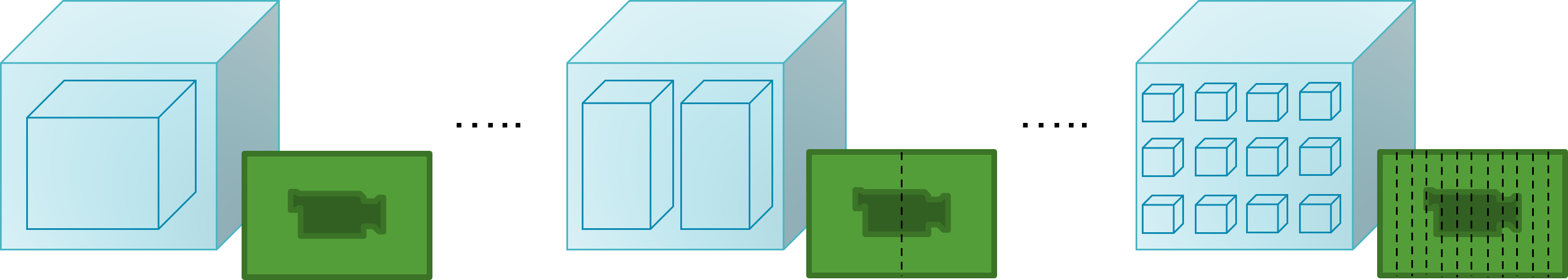}}
\caption{Different experiment scenarios for Jetson TX2 (green) and Jetson AGX Orin (blue) devices. The processing units are equally split between the containers in each device. The input video is also split into equal segments along the temporal dimension, resulting in each container processing an equal number of frames.}
\label{2contain}
\end{figure}

To show the impact of utilizing multiple containers in parallel, we conducted experiments by varying the number of segmented sections and corresponding containers. Then, the inference time and the energy drained in each scenario were recorded. We evaluated the performance of the proposed method on both Jetson TX2 and AGX Orin devices and we repeated the experiments multiple times to ensure the accuracy of the results. As the confidence interval for the measured data of the multiple runs is less than $1\%$, we have opted not to display it in the plots. The number of containers and split segments used in our experiments was limited by the memory capacity and processing power of each edge device, with a maximum of six containers on the Jetson TX2 and twelve containers on the AGX Orin. Fig.~\ref{2contain} depicts the process.

\section{Empirical results and analysis}
\label{sec:res}
In the following plots, as a {\it benchmark}, we consider the case of a single container running YOLO, i.e., a single container, with no data splitting, and all CPU cores (four CPU cores for TX2 and twelve for Orin). All performance metrics (average power, energy and processing time) are normalized with respect to those in the benchmark scenario.  

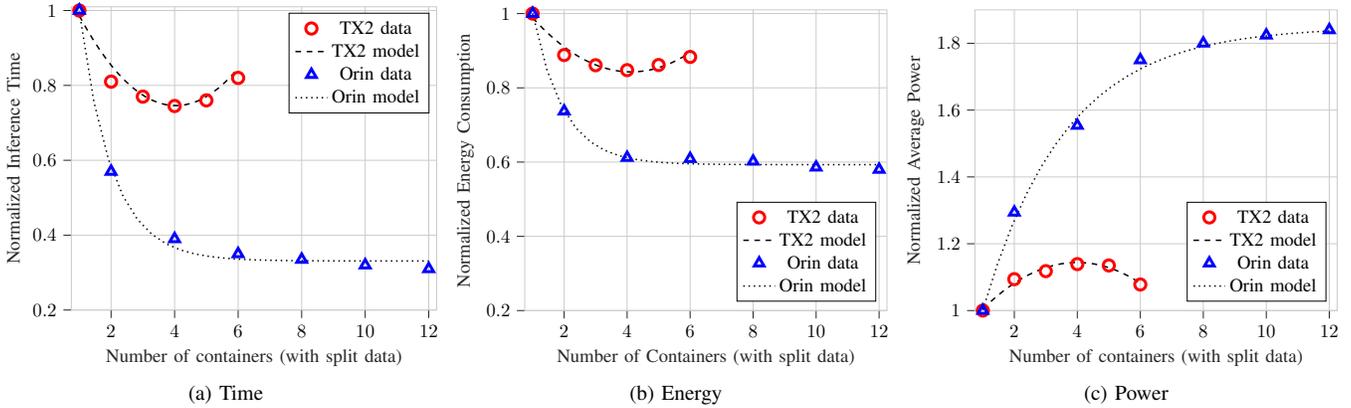
\begin{figure*}
    \centering
    \subfloat[Time\label{fig:time}]{\resizebox{.33\textwidth}{!}{\begin{tikzpicture}

\definecolor{darkslategray38}{RGB}{38,38,38}
\definecolor{lightcoral}{RGB}{240,128,128}
\definecolor{mediumorchid}{RGB}{186,85,211}
\definecolor{lightgray204}{RGB}{204,204,204}

\begin{axis}[
axis line style={lightgray204},
tick align=outside,
tick pos=left,
x grid style={lightgray204},
xlabel=\textcolor{darkslategray38}{Number of containers (with split data)},
xmajorgrids,
xmin=0.75, xmax=12.25,
xtick style={color=darkslategray38},
y grid style={lightgray204},
ylabel=\textcolor{darkslategray38}{Normalized Inference Time},
ymajorgrids,
yminorgrids,
ymin=0.2, ymax=1.01,
ytick style={color=darkslategray38}
]
\addplot [only marks, ultra thick, red, mark=o, mark size= 3pt]
table {%
1 1
2 0.81
3 0.77
4 0.745
5 0.76
6 0.82
};
\addlegendentry{TX2 data}
\addplot[domain=1:6, dashed, thick] {0.026 * x^2 - 0.21 * x + 1.17};
\addlegendentry{TX2 model}

\addplot [only marks, ultra thick, blue, mark=triangle, mark size= 3 pt]
table {%
1 1
2 0.57
4 0.39
6 0.35
8 0.335
10 0.32
12 0.31
};
\addlegendentry{Orin data}
\addplot[domain=1:12, dotted, thick] {1.772 * exp{-0.9786*x + 0.331}}; 
\addlegendentry{Orin model}
\end{axis}

\end{tikzpicture}}}
    \subfloat[Energy\label{fig:energy}]{\resizebox{.33\textwidth}{!}{\begin{tikzpicture}

\definecolor{darkslategray38}{RGB}{38,38,38}
\definecolor{lightgray204}{RGB}{204,204,204}
\definecolor{mediumorchid}{RGB}{186,85,211}

\begin{axis}[
legend pos = south east,
axis line style={lightgray204},
tick align=outside,
tick pos=left,
x grid style={lightgray204},
xlabel=\textcolor{darkslategray38}{Number of Containers (with split data)},
xmajorgrids,
xmin=0.75, xmax=12.25,
xtick style={color=darkslategray38},
y grid style={lightgray204},
ylabel=\textcolor{darkslategray38}{Normalized Energy Consumption},
ymajorgrids,
ymin=0.2, ymax=1.01,
ytick style={color=darkslategray38}
]
\addplot [only marks, ultra thick, red, mark=o, mark size= 3pt]
table {%
1 1
2 0.8883
3 0.8609
4 0.8477
5 0.8615
6 0.8832
};
\addlegendentry{TX2 data}
\addplot[domain=1:6, dashed, thick] {0.0149 * x^2 - 0.123 * x + 1.0967};
\addlegendentry{TX2 model}

\addplot [only marks, ultra thick, blue, mark=triangle, mark size= 3 pt]
table {%
1 1
2 0.737
4 0.612
6 0.609
8 0.602
10 0.586
12 0.58
};
\addlegendentry{Orin data}
\addplot[domain=1:12, dotted, thick] {1.141 * exp{-1.03*x + 0.593}}; 
\addlegendentry{Orin model}
\end{axis}

\end{tikzpicture}}}
    \subfloat[Power\label{fig:power}]{\resizebox{.33\textwidth}{!}{
\begin{tikzpicture}

\definecolor{darkslategray38}{RGB}{38,38,38}
\definecolor{lightcoral}{RGB}{240,128,128}
\definecolor{mediumorchid}{RGB}{186,85,211}
\definecolor{lightgray204}{RGB}{204,204,204}

\begin{axis}[
legend pos = south east,
axis line style={lightgray204},
tick align=outside,
tick pos=left,
x grid style={lightgray204},
xlabel=\textcolor{darkslategray38}{Number of containers (with split data)},
xmajorgrids,
xmin=0.75, xmax=12.25,
xtick style={color=darkslategray38},
y grid style={lightgray204},
ylabel=\textcolor{darkslategray38}{Normalized Average Power},
ymajorgrids,
yminorgrids,
ymin=1, ymax=1.9,
ytick style={color=darkslategray38}
]
\addplot [only marks, ultra thick, red, mark=o, mark size= 3pt]
table {%
1 1
2 1.094
3 1.118
4 1.139
5 1.135
6 1.078
};

\addlegendentry{TX2 data}
\addplot[domain=1:6, dashed, thick] {- 0.0155 * x^2 +0.124 * x + 0.896};
\addlegendentry{TX2 model}

\addplot [only marks, ultra thick, blue, mark=triangle, mark size= 3 pt]
table {%
1 1
2 1.294
4 1.554
6 1.75
8 1.8
10 1.824
12 1.84
};
\addlegendentry{Orin data}
\addplot[domain=1:12, dotted, thick] {-1.24 * exp{-0.38*x} + 1.85}; 
\addlegendentry{Orin model}
\end{axis}

\end{tikzpicture}}}
    \caption{Normalized energy consumption, computation time and average power to execute the object detection task on Jetson TX2 and AGX Orin devices in different scenarios. In every scenario, we use a specific number of containers, each having an equal amount of divided data and processing units.}
    \label{fig3}
\end{figure*}

Fig.~\ref{fig:time} and Fig.~\ref{fig:energy} respectively show the normalized {\it inference time} and {\it energy consumed} for an increasing number of containers used to execute the object detection task, as explained in Section~\ref{sec:meth}. For the Jetson TX2, when running the task on two containers, each using two of the four cores and processing half of the video frames in parallel, we observe a $19\%$ reduction in the inference time (Fig.\ref{fig:time}) and a $10\%$ reduction in the energy consumption (Fig.~\ref{fig:energy}) compared to the benchmark. Increasing the number of containers to four reduced the time further by $25\%$ and the energy by $15\%$. As the number of containers increases beyond four, the system performance degrades in terms of both time and energy. We believe that when the number of containers is increased beyond the number of available CPU cores on the Jetson TX2, it becomes challenging for the CPU scheduler to allocate the CPU cores effectively, worsening the performance. 



On the AGX Orin, which is a more powerful and energy-efficient edge device, splitting the computations between two and four containers respectively results in reductions of $43\%$ and $62\%$ in inference time, and in $25\%$ and $40\%$ reductions in energy consumption, as compared to utilizing a single container. Additionally, increasing the number of containers to twelve on the Orin results in the most efficient scenario for the considered video object detection task. This leads to reductions of about $70\%$ in the inference time and $43\%$ in the energy consumption. However, time and energy curves flatten beyond four containers: since memory resources are used to open new containers, limiting to four can be a good choice.



By analyzing the average power of the devices in each scenario, we get a deeper understanding of how the processing resources are employed in each case. Fig.~\ref{fig:power} shows the average power for an increasing number of containers. We see that splitting the resources and the data among multiple containers leads to a power increase. For the TX2, from one to four containers, we measured a~$13\%$ increase in the average power, while, for the AGX Orin, the increase is about~$84\%$ for twelve containers. This increased average power reflects a better utilization of the available processing resources (CPU cores) on the edge device, which translates into a higher energy efficiency.

Model fitting is also performed for each of the performance metrics (time, power, and energy). The inferred formulae are given in Table~\ref{models}. These fitted models (Fig.~\ref{fig3}) can be beneficial for estimating the savings when applying our method, based on a given reference value (``Ref.'' in Tab.~\ref{models}). In our case the reference value corresponds to the metrics for the benchmark scenario.

We believe that presenting energy and time metrics, in addition to the average power for the different scenarios, provides valuable insights for understanding fundamental \mbox{trade-offs} in edge computing systems. These \mbox{trade-offs} are particularly useful in cases where there are constraints on power or energy usage for the devices. It is also worth noting that the approach outlined in this paper can be extended to other similar tasks and models, not just the video object detection tasks using YOLO-v4Tiny as used in our experiments. We also applied the proposed splitting method to a simple CNN inference task. Splitting the input data (images) between containers led to similar improvements.



\begin{table*}
\centering
\caption{Reference values and fitted models ($x$ means number of containers).}
\label{tab:model_parameters}
\begin{tabularx}{.6\textwidth}{lcYcY}
\toprule
       & \multicolumn{2}{c}{\textbf{TX2}}\quad\quad\quad\quad\quad& \multicolumn{2}{c}{\textbf{AGX Orin}}\quad\quad\quad\quad\quad\\ 
       & \textbf{Ref.} & \textbf{Model} & \textbf{Ref.} & \textbf{Model} \\ \midrule\midrule
\textbf{Time}   &    325 s          &   $0.026 x^2 - 0.21 x + 1.17$        &  54 s                  & $0.33 + 1.77e^{-0.98 x}$               \\
\textbf{Energy} &   942 J            &      $0.015 x^2 - 0.12 x + 1.10$     &              700 J      &   $0.59 + 1.14 e^{-1.03x} $          \\ 
\textbf{Power}  &       2.9 W        &    $- 0.016 x^2 +0.12 x + 0.90 $      &              13 W      &         $ 1.85 -1.24 e^{-0.38x} $      \\ \bottomrule
\end{tabularx}
\label{models}
\end{table*}

\section{Conclusion and future works}
\label{sec:concl}
In this paper, we have proposed a simple methodology to distribute the workload among multiple containers within an edge device. Experiments were conducted on Nvidia Jetson TX2 and AGX Orin boards to analyze the effects on energy consumption and computational time of splitting data and assigning them to multiple containers, while always using all the CPU cores available. Our experimental results show that this procedure yields significant reductions in energy consumption and computational time for an inference task on both devices. 
Notably, this approach is simple and easy to implement, making it a practical solution for edge computing systems employing embedded \mbox{limited-memory} devices.

We remark that the data for the object detection task that was considered in this paper could easily be split into a number of segments by neither negatively impacting the performance nor the accuracy of the model's inference. This descends from the way in which YOLO works, i.e., processing video frames {\it independently} of one another. Our presented approach is thus only applicable when the task can be split into independent subtasks. 
One possible direction for future work could be to investigate the applicability of our method to other types of tasks and models, such as tasks involving data with time correlation or that can only be split into {\it dependent} subtasks. Additionally, it would be interesting to explore the use of our splitting approach in a distributed edge computing setting, where multiple devices collaborate to perform a task. Finally, our method, as well as the results presented in this paper, can be used in the design of \mbox{energy-efficient} job schedulers that split input data, obtaining the optimal number of containers in an online fashion in order to enhance the energy efficiency and reduce the processing time of the edge system.

\bibliographystyle{IEEEtran}
\bibliography{Ref}

\begin{thebibliography}{10}
\providecommand{\url}[1]{#1}
\csname url@samestyle\endcsname
\providecommand{\newblock}{\relax}
\providecommand{\bibinfo}[2]{#2}
\providecommand{\BIBentrySTDinterwordspacing}{\spaceskip=0pt\relax}
\providecommand{\BIBentryALTinterwordstretchfactor}{4}
\providecommand{\BIBentryALTinterwordspacing}{\spaceskip=\fontdimen2\font plus
\BIBentryALTinterwordstretchfactor\fontdimen3\font minus
  \fontdimen4\font\relax}
\providecommand{\BIBforeignlanguage}[2]{{%
\expandafter\ifx\csname l@#1\endcsname\relax
\typeout{** WARNING: IEEEtran.bst: No hyphenation pattern has been}%
\typeout{** loaded for the language `#1'. Using the pattern for}%
\typeout{** the default language instead.}%
\else
\language=\csname l@#1\endcsname
\fi
#2}}
\providecommand{\BIBdecl}{\relax}
\BIBdecl

\bibitem{yolo}
J.~Redmon, S.~Divvala, R.~Girshick, and A.~Farhadi, ``You only look once:
  Unified, real-time object detection,'' in \emph{2016 IEEE Conference on
  Computer Vision and Pattern Recognition (CVPR)}, 2016, pp. 779--788.

\bibitem{yolov4Tiny}
\BIBentryALTinterwordspacing
Z.~Jiang, L.~Zhao, S.~Li, and Y.~Jia, ``Real-time object detection method based
  on improved yolov4-tiny,'' \emph{CoRR}, vol. abs/2011.04244, 2020. [Online].
  Available: \url{https://arxiv.org/abs/2011.04244}
\BIBentrySTDinterwordspacing

\bibitem{oleghe2021container}
O.~Oleghe, ``Container placement and migration in edge computing: Concept and
  scheduling models,'' \emph{IEEE Access}, vol.~9, pp. 68\,028--68\,043, 2021.

\bibitem{matsubara2022split}
\BIBentryALTinterwordspacing
Y.~Matsubara, M.~Levorato, and F.~Restuccia, ``Split computing and early
  exiting for deep learning applications: Survey and research challenges,''
  \emph{ACM Comput. Surv.}, vol.~55, no.~5, dec 2022. [Online]. Available:
  \url{https://doi.org/10.1145/3527155}
\BIBentrySTDinterwordspacing

\bibitem{rodrigues2017fine}
C.~F. Rodrigues, G.~Riley, and M.~Luján, ``Fine-grained energy profiling for
  deep convolutional neural networks on the jetson tx1,'' in \emph{2017 IEEE
  International Symposium on Workload Characterization (IISWC)}, 2017, pp.
  114--115.

\bibitem{suzen2020benchmark}
A.~A. Süzen, B.~Duman, and B.~Şen, ``Benchmark analysis of jetson tx2, jetson
  nano and raspberry pi using deep-cnn,'' in \emph{2020 International Congress
  on Human-Computer Interaction, Optimization and Robotic Applications (HORA)},
  2020, pp. 1--5.

\bibitem{holly2020profiling}
S.~Holly, A.~Wendt, and M.~Lechner, ``Profiling energy consumption of deep
  neural networks on nvidia jetson nano,'' in \emph{2020 11th International
  Green and Sustainable Computing Workshops (IGSC)}, 2020, pp. 1--6.

\bibitem{lahmer2022energy}
S.~Lahmer, A.~Khoshsirat, M.~Rossi, and A.~Zanella, ``Energy consumption of
  neural networks on nvidia edge boards: an empirical model,'' in \emph{2022
  20th International Symposium on Modeling and Optimization in Mobile, Ad hoc,
  and Wireless Networks (WiOpt)}, 2022, pp. 365--371.

\bibitem{dockkvm}
B.~Wang, J.~Xie, S.~Li, Y.~Wan, S.~Fu, and K.~Lu, ``Enabling high-performance
  onboard computing with virtualization for unmanned aerial systems,'' in
  \emph{2018 International Conference on Unmanned Aircraft Systems (ICUAS)},
  2018, pp. 202--211.

\bibitem{shalavi2022energy}
N.~Shalavi, G.~Perin, A.~Zanella, and M.~Rossi, ``Energy efficient deployment
  and orchestration of computing resources at the network edge: a survey on
  algorithms, trends and open challenges,'' \emph{arXiv preprint
  arXiv:2209.14141}, 2022.

\bibitem{kaur2020keids}
K.~Kaur, S.~Garg, G.~Kaddoum, S.~H. Ahmed, and M.~Atiquzzaman, ``Keids:
  Kubernetes-based energy and interference driven scheduler for industrial iot
  in edge-cloud ecosystem,'' \emph{IEEE Internet of Things Journal}, vol.~7,
  no.~5, pp. 4228--4237, 2020.

\bibitem{kubernetes}
\BIBentryALTinterwordspacing
Google. Kubernetes. [Online]. Available: \url{https://kubernetes.io/}
\BIBentrySTDinterwordspacing

\bibitem{perin2022ease}
G.~Perin, F.~Meneghello, R.~Carli, L.~Schenato, and M.~Rossi, ``{EASE:
  Energy-Aware Job Scheduling for Vehicular Edge Networks With Renewable Energy
  Resources},'' \emph{IEEE Transactions on Green Communications and
  Networking}, pp. 1--1, 2022.

\bibitem{Yolov4}
\BIBentryALTinterwordspacing
A.~Bochkovskiy, C.-Y. Wang, and H.-Y.~M. Liao, ``Yolov4: Optimal speed and
  accuracy of object detection,'' 2020. [Online]. Available:
  \url{https://arxiv.org/abs/2004.10934}
\BIBentrySTDinterwordspacing

\bibitem{docker}
D.~Merkel, ``Docker: lightweight linux containers for consistent development
  and deployment,'' \emph{Linux journal}, vol. 2014, no. 239, p.~2, 2014.

\end{thebibliography}


\end{document}